\pdfoutput=1
\documentclass[aip,reprint]{revtex4-2}

\usepackage[pdftex]{graphicx}
\usepackage[pdftex]{epsfig}
\usepackage{epstopdf}
\usepackage[dvipsnames]{xcolor}
\usepackage{amssymb,amsmath}
\usepackage{graphicx}
\usepackage{dcolumn}
\usepackage{multirow}
\usepackage{hyperref}
\hypersetup{colorlinks,urlcolor=blue,citecolor=blue}
\usepackage{siunitx}

\DeclareMathOperator{\imag}{Im}

\begin{document}

\title{Improving dynamic collision frequencies: impacts on dynamic structure factors and stopping powers in warm dense matter}

\author{Thomas W. Hentschel}
\affiliation{School of Applied \& Engineering Physics, Cornell University, Ithaca NY, USA \looseness=-1}
\author{Alina Kononov}
\affiliation{Center for Computing Research, Sandia National Laboratories, Albuquerque NM, USA \looseness=-1}
\author{Alexandra Olmstead}
\affiliation{Center for Computing Research, Sandia National Laboratories, Albuquerque NM, USA  \looseness=-1}
\affiliation{Nanoscience and Microsystems Engineering Program, University of New Mexico, Albuquerque, NM, USA  \looseness=-1}
\author{Attila Cangi}
\affiliation{Center for Advanced Systems Understanding, Helmholtz-Zentrum Dresden-Rossendorf, G\"orlitz, Germany \looseness=-1}
\author{Andrew D. Baczewski}
\affiliation{Center for Computing Research, Sandia National Laboratories, Albuquerque NM, USA \looseness=-1}
\affiliation{Center for Quantum Information and Control (CQuIC), Department of Physics and Astronomy, University of New Mexico, Albuquerque, NM, USA \looseness=-1}
\author{Stephanie B. Hansen}
\affiliation{Pulsed Power Sciences Center, Sandia National Laboratories, Albuquerque NM, USA \looseness=-1}

\begin{abstract}
Simulations and diagnostics of high-energy-density plasmas and warm dense matter 
rely on models of material response properties, both static and dynamic (frequency-dependent). Here, we systematically investigate variations in dynamic electron-ion collision frequencies $\nu(\omega)$ in warm dense matter using data from a self-consistent-field average-atom model. We show that including the full quantum density of states, strong collisions, and inelastic collisions lead to significant changes in $\nu(\omega)$. These changes result in red shifts and broadening of the plasmon peak in the dynamic structure factor, an effect observable in x-ray Thomson scattering spectra, and modify stopping powers around the Bragg peak. These changes improve the agreement of computationally efficient average-atom models with first-principles time-dependent density functional theory in warm dense aluminum, carbon, and deuterium.
\end{abstract}

\maketitle

\section{Introduction} 
\label{sec:introduction}
In high-energy-density plasmas such as those occurring in inertial confinement fusion (ICF) implosions, stellar atmospheres, and planetary interiors, atomic-scale collisions between electrons and ions have a cascading impact at longer length and time scales: collisions mediate transport properties, dampen collective plasma oscillations, and ultimately contribute to the particle and energy transport processes that govern plasma evolution. 

In the static (zero-frequency) limit, collision frequencies inform properties like electrical and thermal conductivities that can impact instability development in ICF implosions \cite{HAMMEL2010171}. Dynamic (frequency-dependent) collisions influence plasma response functions that influence measurable quantities like x-ray Thomson Scattering (XRTS) spectra \cite{GlenzerRedmer2009} and stopping powers \cite{Zylstra2015,malko2022proton}. Stopping powers are especially important for ICF since they mediate the velocity-dependent return of energy from fast fusion products back into the fusion fuel and play a critical role in self-heating and ignition \cite{Lindl1995, Hurricane2016, zylstra2019alpha}.

While there is no direct observable associated with dynamic collision frequencies, their static limit is closely tied to measurable electrical conductivities \cite{milchberg1988resistivity, desilva}, and their dynamic behavior can be constrained with x-ray Thomson Scattering (XRTS) measurements \cite{Sperling2015}. However, direct, benchmark-quality measurements of these transport and observable properties are difficult because they require preparing, independently characterizing, and interrogating uniform high-energy-density samples \cite{frenje2015measurements, Zylstra2015, malko2022proton}. Calculating these properties is also challenging, particularly in the warm dense matter (WDM) regime (materials near solid densities with temperatures near the Fermi energy) since it requires accounting for many different physical effects including partial ionization, screening, degeneracy, and ion coupling. Many models of collisional processes in use today rely on approximations drawn from plasma physics (e.g. weak scattering, ideal electron distributions, and Coulomb logarithms \cite{Brysk1974}) that are known to be inaccurate for WDM.

Fortunately, sophisticated first-principles models have been developed that can self-consistently follow the complete electronic structure and ionic motion of 10s-100s of atoms in a WDM state. These models are based on density functional theory (DFT)~\cite{hohenberg1964inhomogeneous, kohn1965self} and provide reliable predictions for equations of state and static transport properties~\cite{desjarlais2002electrical,schorner2022ab}. Additionally, real-time time-dependent DFT (TDDFT)~\cite{runge1984density,marques2006time,ullrich2011time,kononov2022electron} can model the dynamic response of materials to incident electromagnetic waves and charged particles, offering direct access to many transport and observable properties including stopping powers \cite{Correa2012,Schleife2015,Shukri2016,Magyar2016,kononov2020pre,kononov2021anomalous} and dynamic structure factors (DSFs) \cite{Baczewski2016,Ramakrishna2021,Baczewski2021}. 

While first-principles TDDFT models can provide accurate predictions for material properties, like conventional DFT, their computational cost grows with temperature in WDM conditions and becomes even more expensive at higher temperatures and for higher Z species. They are thus generally unsuitable for building the extensive tables required for simulations of many astrophysical and laboratory plasmas. Instead, researchers often rely on simpler models, such as DFT-based average atom (AA) models~\cite{Liberman1979, Wilson2006, Starrett2013, callow2022first}, to populate wide-ranging tables of plasma properties. These simpler models are not only less accurate than multi-atom (TD)DFT, but typically also require external models like the Ziman formula \cite{Ziman,Burrill2016,PainWetta2020} for calculations of static collision frequencies or the random phase approximation (RPA) to account for dynamic effects in scattering \cite{gregori2003theoretical, GlenzerRedmer2009, Johnson2012} and stopping powers \cite{Wang1998, Faussurier2010}. 

DFT-AA models can go beyond the RPA by including static collisions and local field corrections \cite{BarrigaCarrasco2008, BarrigaCarrasco2009, Moldabekov2020} or dynamic collision frequencies $\nu(\omega)$ in their calculations of dynamic response functions \cite{Sperling2015, Souza2014, dharma2016dynamic} using the Mermin modification to the RPA \cite{Mermin1970}. However, the accuracy of these corrections has not been rigorously assessed. Further, most such implementations rely on approximations such as the uniform electron (UEG) gas \cite{Lindhard1964, Arista-Brandt1981, Maynard1982, Skupsky1977, Arista-Piriz1987} or Born collision cross sections \cite{Sperling2015, Souza2014, DWD2016}, neither of which is valid in the WDM regime: the UEG does not capture structure in the continuum electron density of states (DOS), and Born cross sections do not capture the strong collisions that dominate scattering processes near the Fermi energy in partially ionized atoms. Corrections beyond the Born approximation have been limited to combining strong collisions calculated using ladder diagrams with dynamic screening via the Gould-DeWitt approach \cite{Sperling2015, Reinholz2000, Thiele2006} or normalizing  Born collision frequencies to match Ziman calculations using T-matrix collision frequencies at $\omega=0$ \cite{DWD2016}.
Finally, few DFT-AA models consider either the full quantum DOS or the impact of inelastic collisions that can lead to rapid damping of plasma excitations.

In Section \ref{sec:AAnu} of this work, we briefly describe a DFT-AA model suitable for exploring the impact of dynamic collision frequencies $\nu(\omega)$ on DSFs and stopping powers. In Section \ref{sec:nu}, we systematically investigate the impacts of four variations on DFT-AA calculations of $\nu(\omega)$: ideal vs.\ quantum DOS, uniform vs.\ structured ions, static (Born) vs.\ dynamic (Lennard-Balescu) screening, and Born (weak) vs.\ T-matrix (strong) cross sections.  We also propose an approximation for inelastic contributions to $\nu(\omega)$ and compare to $\nu(\omega)$ derived from Kubo-Greenwood calculations of the dynamic conductivity \cite{Reinholz2000}. We find that using the quantum DOS is critical to preserve sum rules, that there is promising agreement between the independent T-matrix and Kubo-Greenwood approaches, that inelastic collisions may have significant impact on $\nu(\omega)$, and that the Born approximation is fundamentally unreliable for WDM. In Section \ref{sec:DSF}, we show that the most complete calculations of $\nu(\omega)$ improve agreement with direct DSF calculations from TDDFT. In Section \ref{sec:dEdx}, we describe our derivation of stopping powers from the Mermin dielectric functions and show that the improvements in the DSFs carry over into significant improvements in stopping powers. We conclude in Section \ref{sec:conclusion}. %

\label{sec:methods}
\section{Average-atom model}

\label{sec:AAnu}

Fully quantum AA models have been used for decades \cite{Liberman1979, Wilson2006, Starrett2013} to describe both thermodynamic and transport properties\cite{Rinker1985,Sterne2007}, and have recently been extended to provide access to direct experimental observables like XRTS spectra \cite{Johnson2012, Souza2014}. The AA model in this manuscript closely follows these previous implementations.

DFT-AA models self-consistently solve a set of Kohn-Sham-like equations. The first step involves guessing an electron-ion potential $V_{ei}(r)$ that assumes a functional form for the exchange-correlation potential $V_{xc}(r)$. We use a $V_{ei}(r)$ that is forced to zero at the Wigner-Seitz radius $R_{WS} = (3/(4\pi n_i))^{1/3}$ for ion density $n_i$ and a local density approximation (LDA) for $V_{xc}(r)$~\cite{Ceperley1980}. Next, we solve the radial Schr\"odinger equation to find bound $P_{n\ell}(r)$ and continuum $P_{\varepsilon\ell}(r)$ radial electronic orbitals, and populate these orbitals according to Fermi-Dirac occupation factors:
\begin{equation}
f(\varepsilon) = \frac{1}{1 + \exp((\varepsilon - \mu)/(k_BT))}
\label{eq:FD-distro}
\end{equation}
with $\varepsilon$ the electron energy and $\mu$ a chemical potential that enforces neutrality within the Wigner-Seitz sphere. The occupied orbitals provide an electron density $n_e(r)$ from $4\pi r^2 n_e(r) = \sum_a f(\varepsilon_a) g_a P_a^2(r)$ with $g_{a} = 2(2\ell_a+1)$ that in turn generates a new $V_{ei}(r)$; this procedure is iterated until $n_e(r)$ and $V_{ei}(r)$ converge to self-consistency. 

At convergence, the bound electrons are represented by orbitals with a fixed number of nodes and energy eigenvalues $\varepsilon_{n\ell}$, while the continuum electrons are represented by distorted waves with phase shifts $\delta_{\varepsilon\ell}$ determined by matching numerical orbitals to analytic plane waves at $R_{WS}$. The continuum orbitals define a quantum density of states $X(\varepsilon)= \sum_{\ell} g_{\ell}\int_0^{R_{WS}}{P_{\varepsilon\ell}^2(r) dr}$, whose accuracy is largely determined by the adequacy of the spherical approximation for $V_{ei}(r)$, the choice of $V_{xc}(r)$, and boundary conditions on $V_{ei}(r)$ and $P_a(r)$ \cite{callow2022first}. 

\begin{figure}
    \centering
    \includegraphics[width=0.45\textwidth]{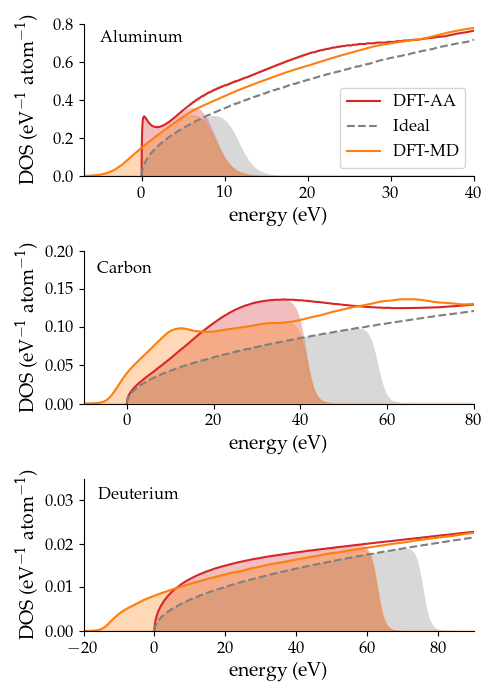}
    \caption{Continuum and ideal free electron DOS for aluminum, carbon, and deuterium at a temperature of $T = \SI{1}{\electronvolt}$ and densities of  $\SI{2.7}{\gram\per\cubic\centi\meter}$ for Al,  $\SI{10}{\gram\per\cubic\centi\meter}$ for C, and $\SI{10}{\gram\per\cubic\centi\meter}$ for D.
    The lightly shaded areas are the product of the Fermi-Dirac distribution with the corresponding DOS, where the chemical potential is chosen in each case such that integral over energy equals 3.0, 4.0, and 1.0 for aluminum, carbon, and deuterium respectively. The energy scales of DFT-MD are shifted to align the chemical potential with that of the DFT-AA model, $\mu_{AA}$.}
    \label{fig:dos}
\end{figure}

Fig.\ \ref{fig:dos} shows the continuum-electron density of states (DOS) $X(\varepsilon)$ obtained from our DFT-AA model (solid red line) at a thermal energy of $\SI{1}{\electronvolt}$ and densities of  $\SI{2.7}{\gram\per\cubic\centi\meter}$ for aluminum,  $\SI{10}{\gram\per\cubic\centi\meter}$ for carbon, and $\SI{10}{\gram\per\cubic\centi\meter}$ for deuterium. The ideal electron gas DOS, $X^i= (2\varepsilon)^{1/2}/(n_i \pi^2)$, with $n_i$ the ion density, is also shown to represent the allowed states for plane-wave (ideal) free electrons. As illustrated here, the ratio of the quantum DOS to the ideal DOS, $\xi(\varepsilon)=X(\varepsilon)/X_i(\varepsilon)$, is typically larger than unity. The occupied states, $X(\varepsilon)f(\varepsilon)$, are shown as the shaded regions; integrals over these regions represent estimates for the screened ion charge $Z_s$. For the fully quantum DOS from DFT-AA, the area contained in the red shaded region gives continuum-electron values $Z_c$ of 3.0, 4.0, and 1.0 for aluminum, carbon, and deuterium respectively, while integrals of the ideal DOS with $f(\varepsilon)$ parameterized by the same chemical potential predicts smaller ideal free-electron values $Z_i$ of 2.0, 2.4, and 0.75 \cite{murillo2013partial}. In order for quantities derived from an ideal density of states to satisfy continuum-electron sum rules, one would have to impose a new chemical potential, $\mu_i$ \cite{Burrill2016}, illustrated by the shaded gray regions. 

We can directly compare these results against first-principles DFT-MD calculations (solid orange lines), which include non-spherical effects and can support continuum states with negative energies. Since DFT-MD does not necessarily use the same energy zero as AA and deuterium lacks a bound state to anchor the energy axis, we align the DFT-MD curves to match the DFT-AA chemical potential. For all elements, the DFT-MD DOS is closer to the DFT-AA quantum DOS than to the ideal DOS. Thermal broadening of the low-energy features would further improve the agreement. For carbon, the DFT-MD DOS has additional structure absent in the the DFT-AA model, suggesting residual non-spherical or band-structure effects.

\begin{figure}
    \centering
    \includegraphics[width=0.45\textwidth]{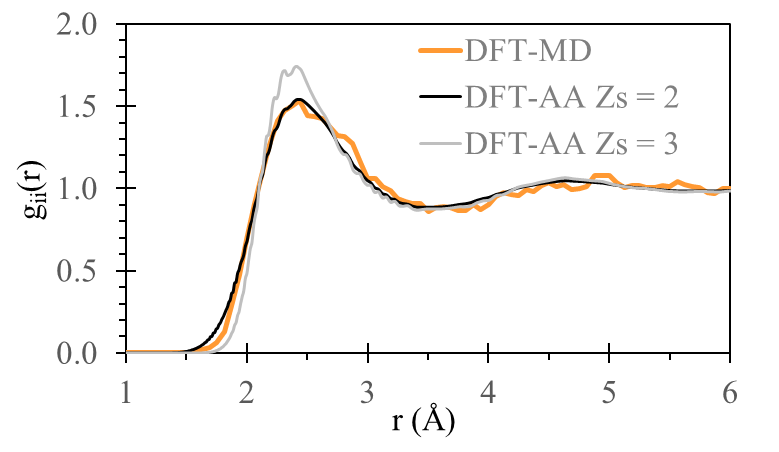}
    \caption{Radial ion distribution functions for solid-density Al at \SI{1}{\electronvolt} from the DFT-AA model with different choices for the screened ion charge $Z_s$, compared with \emph{ab-initio} DFT-MD results. This distribution is directly related to the ion-ion structure factor $S_{ii}(k)$. 
    }
    \label{fig:gofr}
\end{figure}

To model ionic properties, we follow the methods of Starrett and Saumon \cite{Starrett2012, Starrett2013}, solving a second set of self-consistent equations to obtain the potential and electron density associated with the external plasma. Once the external density is determined, one must assign a screened ion charge $Z_s$ to determine the ion-ion potential $V_{ii}(r)$ and generate a static ion-ion structure factor $S_{ii}(k)$ and its Fourier transform, the radial ion distribution function $g_{ii}(r)$. We have found consistently good agreement with \emph{ab-initio} multi-center DFT-MD models using $Z_s = Z_i$. For example, Fig.\ \ref{fig:gofr} shows that a DFT-AA model using $Z_s = Z_i = 2$ for aluminum with a thermal energy of \SI{1}{\electronvolt} gives better agreement with DFT-MD calculations of $g_{ii}(r)$ than one that uses $Z_s = Z_c = 3$. In contrast to quantities  that integrate over the DOS, like the collision frequencies, DSFs, and stopping powers described in the following sections, here the choice of $Z_s$ is unconstrained by sum rules. %

\section{Electron-ion collision frequencies}
\label{sec:nu}

The DFT-AA model described above provides an internally consistent picture of both electronic and ionic structure. In principle, it provides sufficient information to compute many thermodynamic and transport quantities, including static collision frequencies $\nu(0)$ in the Ziman formulation \cite{Ziman}:
\begin{equation}
\nu^Z(0) = \frac{1}{3\pi Z_0} \int_0^\infty d\varepsilon \int_0^{2p} dk \, k^3 S_{ii}(k) 
 \frac{\partial f(\varepsilon)}{\partial \varepsilon} \frac{\partial \Sigma(\varepsilon,\theta)}{\partial \theta}
\label{eq:Ziman}
\end{equation}
and dynamic collision frequencies $\nu(\omega)$ from the generalized Boltzmann equation \cite{Thiele2006, Thiele2008, Souza2014, faussurier2016electron}:
\begin{equation}
\nu(\omega) = \frac{-i}{6\pi Z_0}
\int_0^{\infty} dk \, k^6 S_{ii}(k) \frac{\partial \Sigma(\varepsilon,\theta)}{\partial \theta}
\frac{\epsilon^{0}(k,\omega)-\epsilon^{0}(k,0)}{\omega}
\label{eq:vw}
\end{equation}

In both of the above expressions, $\frac{\partial \Sigma}{\partial \theta}$ is a differential cross section that represents the  magnitude of momentum transfer for particles with momentum $p=(2\varepsilon)^{1/2}$ scattering at a given angle $\theta$, with $k^2 = 2p^2(1-\cos\theta)$ and $S_{ii}(k)$ the static ion-ion structure factor. In Eq.\ \eqref{eq:Ziman}, the normalization factor $Z_0$ is constrained through a finite-temperature generalization of Fermi surface properties \cite{RinkerPRA88}. In the UEG (ideal) gas approximation implicit in Eq.\ \eqref{eq:Ziman}, $Z_0=Z_i$. In \mbox{Eq.\ \eqref{eq:vw}}, $\epsilon^{0}(k,\omega)$ is the RPA dielectric function, which is derived from a UEG approximation that also implicitly assumes an ideal electron gas DOS, for which we set $Z_0=Z_i$.

Both of these expressions are highly integrated quantities with multiple dependencies, and both affect observable transport and response properties. The static collision frequency directly informs the electrical conductivity $\sigma_{DC}= Z_s n_i/\nu(0)$ and the dynamic collision frequency $\nu(\omega)$ influences DSFs and stopping powers (see Sections \ref{sec:DSF} and \ref{sec:dEdx}). In the remainder of this section, we systematically explore the impact of variations in $\frac{\partial \Sigma}{\partial \theta}$, $S_{ii}(k)$, DOS, and static and dynamic screening. %

\subsection{Electron-ion collision cross sections}
\label{ssec:eicc}

Electron-ion collision cross sections describe the interaction of an impact electron at energy $\varepsilon$ with an ion defined by a screened nuclear charge. A very common simplification is the assumption of weak collisions, which permits use of the Born approximation. For a given scattering potential $V^s(r)$, the Born approximation is defined by the Fourier transform $\Tilde{V}^s(k)$ \cite{Griffiths, Sakurai}:
\begin{equation}
\frac{\partial \Sigma^B(\varepsilon,\theta)}{\partial \theta}= \frac{| \Tilde{V} ^s (k)|^2 } {4\pi^2}
\label{eq:sigB}
\end{equation}

\begin{figure}
    \centering
    \includegraphics[width=0.45\textwidth]{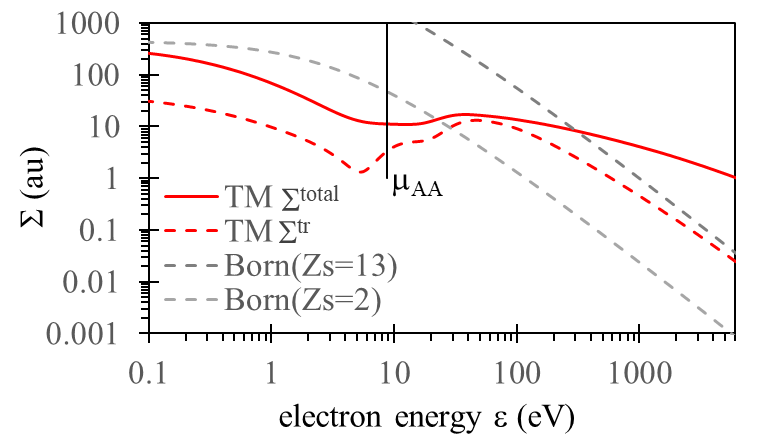}
    \caption{Scattering cross sections for solid-density aluminum at \SI{1}{\electronvolt} from Eqs.\ \eqref{eq:sigB} and \eqref{eq:sigT} integrated over the solid angle. Dashed gray lines show Born-Yukawa approximations with two different choices of ion charge $Z_s$. T-matrix results from the DFT-AA model are given for the momentum-transfer (dashed red, Eq.\ \eqref{eq:sigtr} with a self-consistent $S_{ii}(k)$) and total (solid red, Eq.\ \eqref{eq:sigt-tot}) cross sections. 
    }
    \label{fig:Sigma}
\end{figure}

There are many plausible choices for the scattering potential, including the self-consistent potentials of the DFT-AA model \cite{Burrill2016}. In general, these self-consistent potentials can be fit to a simple parameterized Yukawa form $V^s(r) = -Z_s e^{-k_sr}/r$, which implies $\Tilde{V}^s(k) = - 4 \pi Z_s / (k^2+k_s^2)$, with $Z_s$ the screened (background) nuclear charge and $k_s$ the inverse of the screening length $r_s = ( T_{\mathrm{eff}}/4\pi n_e)^{1/2}$. Here, $T_{\mathrm{eff}} = (T^2+T_F^2)^{1/2}$ is an effective temperature that interpolates between classical and degenerate cases, with $T_F$ the Fermi energy. Examples of Born-Yukawa cross sections for aluminum at solid density and $T =$ \SI{1}{\electronvolt} with two different choices of $Z_s$ are given in Figure\ \ref{fig:Sigma}. Here, we show $Z_s = 2$ to represent scattering from the screened nucleus with $Z_i$ free electrons and and $Z_s = 13$ to represent scattering from the unscreened nucleus: this choice uniquely determines the high-energy behavior. 

The Born approximation is only accurate for weak scattering from a potential that minimally perturbs the free-electron waves; \emph{i.e.} for incident electrons with energies greater than the magnitude of the scattering potential \cite{Griffiths}. The screening constant $k_s$ softens the scattering interaction for low-energy collisions, but it cannot capture the complexity of fully quantum screening in partially ionized materials. 

To account for strong scattering in the self-consistent potential of the DFT-AA model, we use a transition- or T-matrix formulation that explicitly includes the effects of the distorted quantum continuum electron orbitals in the presence of the self-consistent $V_{ei}$ through their phase shifts $\delta_{\varepsilon\ell}$ \cite{Burrill2016, Sakurai}:
\begin{equation}
\frac{\partial \Sigma^T(\varepsilon,\theta)}{\partial \theta}= \frac{1}{p^2} \left| \sum_{\ell=0}^{\infty} (2\ell+1) \sin \delta_{\varepsilon\ell} \,e^{i\delta_{\varepsilon\ell}}P_{\ell} (\cos\theta) \right| ^2,
\label{eq:sigT}
\end{equation}
where $P_{\ell}$ are Legendre polynomials. For $S_{ii}(k) = 1$, the momentum-transfer T-matrix cross section can be integrated over the solid angle to give the energy-dependent cross section: 
\begin{equation}
\Sigma^{\mathrm{tr}}(\varepsilon) = \frac{4\pi}{p^2} \sum_{\ell=0}^{\infty} (\ell+1) \sin^2 (\delta_{\varepsilon\ell}-\delta_{\varepsilon\ell+1}).
\label{eq:sigtr}
\end{equation}

At high energies, this expression approaches the $Z_s=13$ limit of the momentum-integrated Born-Yukawa cross section, 
$\Sigma^{tr}(\varepsilon)= \pi Z_s^{2}/(2\varepsilon^2)(x-\text{ln}(x)-1)$ with $x=k_s^2/(k_s^2+8\varepsilon)$. At lower energies, the momentum-integrated Born and T-matrix cross sections have very different behavior, including near the $\mu_{AA} \approx E_F \approx$ \SI{10}{\electronvolt} electron impact energies that dominate thermal scattering. While the Born approximation with $Z_s=2$ is more reasonable at these lower energies than the Born approximation with $Z_s=13$, it is not generally reliable for WDM.

Figure\ \ref{fig:Sigma} also shows the total cross section, 
\begin{equation}
\Sigma^{\mathrm{total}}(\varepsilon) = \frac{4\pi}{p^2} \sum_{\ell=0}^{\infty} (2\ell+1) \sin^2 (\delta_{\varepsilon\ell}),
\label{eq:sigt-tot}
\end{equation}
which we will use below to estimate inelastic contributions to $\nu(\omega)$.

\subsection{$\nu(\omega)$ from electron-ion collision integrals}
\label{ssec:eiint}

\begin{figure*}
    \centering
    \includegraphics[width=0.9\textwidth]{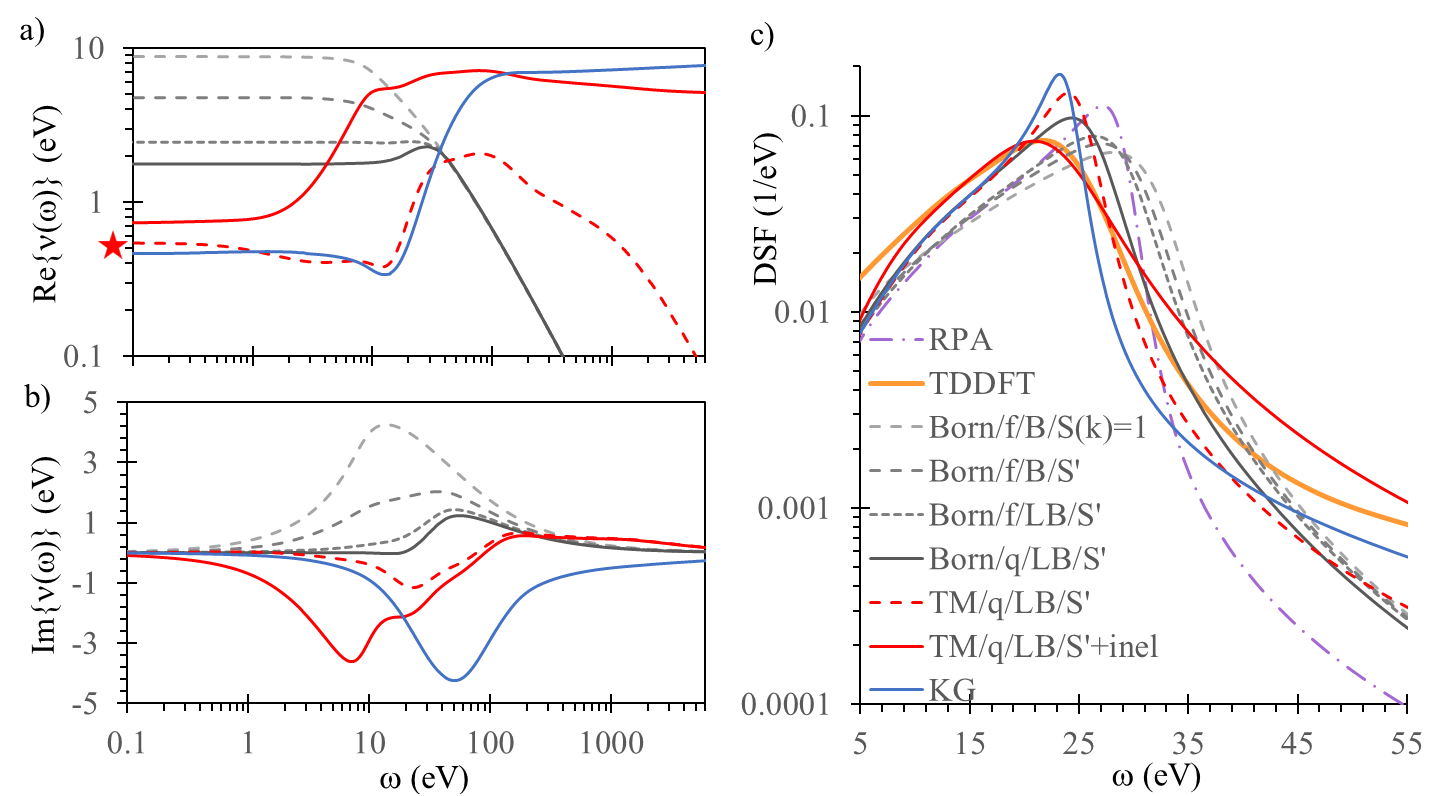}
    \caption{Real (a) and imaginary (b) parts of the dynamic collision frequency of solid-density aluminum at T = \SI{1}{\electronvolt} from various approximations described in the text. The red star indicates a reference point for the static collision frequency
    \cite{milchberg1988resistivity,starrett2020tabular}. (c) Mermin dynamic structure factors at a wavenumber of \SI{1.55}{\per\angstrom} corresponding to the various $\nu(\omega)$ described in the text.
    The orange curve from TDDFT is the reference spectrum, similar to experimental data at similar conditions presented in~\citet{Sperling2015}. In the legend, Born/TM indicates the cross section; f/q indicates the DOS; B/LB indicates Boltzmann or Lenard-Balescu integration; and S' indicates a self-consistent structure factor.
    }
    \label{fig:various}
\end{figure*}

The collision frequencies in Eqs.\ \eqref{eq:Ziman} and \eqref{eq:vw} depend not only on collision cross sections but also on the ion-ion structure factors $S_{ii}(k)$ and on the densities of states implied by the $\partial f(\varepsilon)/\partial \varepsilon$ and $\epsilon^0(k,\omega)$ terms in their respective integrals. Here, we systematically investigate changes in these terms, moving from the simplest approximation of a Born-Yukawa cross section with $S_{ii}(k)=1$ and an ideal DOS to a T-matrix cross section with a self-consistent $S_{ii}(k)$ and a quantum density of states. In the process, we will show that moving towards the reference $S_{ii}(k)$ and DOS established above through comparison to DFT-MD improves agreement with reference values for $\nu(0)$ and the dynamic structure factor (DSF).

 Figure \ref{fig:various} shows the real and imaginary parts of $\nu(\omega)$ and its impact on the DSF of solid aluminum at a temperature of \SI{1}{\electronvolt}. In general, the real part of $\nu(\omega)$ will broaden the DSF from its RPA limit and the imaginary part will shift it \cite{Fortmann2010}. Experimental data at similar conditions \cite{Sperling2015} falls close to the orange curve from our reference TDDFT model. Thus a $\nu(\omega)$ with a fairly large positive real value and a fairly large negative imaginary value is needed to broaden and shift the RPA DSF towards the reference spectrum. The real part of $\nu(\omega)$ is also constrained in the static limit by reference conductivity data \cite{milchberg1988resistivity,starrett2020tabular}; this static collision frequency is indicated by the red star.

The simplest approach to calculating $\nu(\omega)$ is given by the light gray dashed lines in Fig.\ \ref{fig:various}. It uses the Born-Yukawa cross section with $Z_s= Z_i$, the ideal free-electron density of states, $S_{ii}(k)=1$, and a direct integration over $\epsilon(k,\omega)$ (denoted ``B'' in the legend). Using the Born cross section simplifies the expressions for both $\nu^Z(0)$ and $\nu(\omega)$, replacing the energy integral in $\nu^Z(0)$ with a factor $f(k/2)$ \cite{Burrill2016}. Note that $f(k/2)$ is also the static limit of $\frac{-i k^3}{2\omega}[\epsilon^{0}(k,\omega)-\epsilon^{0}(k,0)]$, which ensures consistency between the two expressions. The real part of this simple $\nu^B(\omega)$ has a static limit that is much larger than the reference value and a fixed functional form that follows an asymptotic decay proportional to $Z_s^{2}/\omega^{3/2}$, turning near the plasma frequency $\omega_{pl} = (4 \pi Z_s n_i)^{1/2}$ of \SI{12.9}{\electronvolt} (for $Z_s= Z_i$). Because of its fixed form, the imaginary part of $\nu(\omega)$ is always positive and so it moves the RPA DSF \emph{away} from the reference spectrum.

The darker dashed gray lines in Fig.\ \ref{fig:various} show the impact of changing $S_{ii}(k)$ from unity to the self-consistent $S_{ii}(k)$ from the DFT-AA model (see Fig.\ \ref{fig:gofr}). Here, we integrate over $\min[S_{ii}(k),1]$ to cut out strong peaks and approximately account for incipient lattice structure in strongly coupled fluids (\emph{c.f.}\ \citet{PainWetta2020,baiko}). This change decreases the static limit but does not change the functional form of the simplest Born approach. 

Next, we modify the integration of Eq. \ref{eq:vw}, substituting $\epsilon^{-1}(k,\omega)$ for $\epsilon(k,\omega)$ and replacing $-i$ with $i$. This Lenard-Balescu (LB) integration better represents dynamic screening \cite{Reinholz2000}. We find results (dotted gray lines) very similar to those of \citet{faussurier2016electron} under this change: a decrease in the magnitude of $\nu(\omega)$ but no change to its functional form. Note that this change removes the formal equivalence between Eq.\ \eqref{eq:Ziman} and the zero-frequency limit of Eq.\ \eqref{eq:vw}. Although $\nu(0)$ from the two expressions are often very close, we normalize low-frequency values to the Ziman limit $\nu^Z(0)$ by setting $\nu(\omega)=[1-f_Z(\omega)]\nu^{LB}(\omega)+f_Z(\omega)\nu^Z(0)$, with $f_Z(\omega) = 1 -[1+(0.1\,\omega_{pl}/\omega)^2]^{-1}$.

Our final calculation with the Born-Yukawa cross section replaces every $f(\varepsilon)$ in the expressions for the RPA $\epsilon(k,\omega)$ \cite{Johnson2012} with $f(\varepsilon)\xi(\varepsilon)$, which replaces the implicitly ideal density of states in the RPA with the fully quantum DOS.
The $\xi(\varepsilon)$ factor is also applied to the Mermin $\epsilon(k,\omega)$ used to obtain the DSF, 
and as a prefactor to the $\partial f(\varepsilon)/\partial \varepsilon$ term in Eq.\ \eqref{eq:Ziman}  \cite{StarrettTab}, which necessitates a change to the normalization factor: $Z_0=Z_c$.  This procedure allows us to use the DFT-AA chemical potential while still accounting for all of the continuum electrons (see details in Appendix \ref{appendix:DOS-change}).

As illustrated by the solid gray lines, using the quantum DOS further reduces the magnitude of $\nu(\omega)$, almost recovering the RPA DSF. It also modifies the shape of $\nu(\omega)$, introducing some functional variation and moving the turning point to higher frequencies consistent with  $\omega_{pl} =$ 15.8 eV for $Z_s= Z_c$. In all Born-Yukawa cases, however, both the static limit of $\nu(\omega)$ and the Mermin DSF are far from the reference values.

Keeping all of the above improvements, we now replace the Born-Yukawa cross section with the T-matrix momentum-transfer cross section given by Eq.\ \eqref{eq:sigtr} and illustrated in in Fig.\ \ref{fig:Sigma}. This change results in the dashed red lines in Fig.\ \ref{fig:various}. With strong collisions, $\nu(\omega)$ takes on a dramatically different functional form whose real part matches the static reference value very well and whose imaginary part moves the RPA peak towards the reference DSF. But while the real part of $\nu(\omega)$ increases from its static limit above the plasma frequency, it is too small near the peak of the DSF to give the necessary broadening.

So far, we have considered only elastic (momentum-transfer) collisions in $\nu(\omega$), which represents the damping of plasma excitations at the driving frequency $\omega$. However, inelastic single-particle excitations of one continuum electron into a higher-energy state are also possible and would be followed by a very rapid recombination that could contribute to the damping of collective modes. To estimate how such inelastic collisions might affect $\nu(\omega$), we define an \emph{ad-hoc} inelastic collision cross section $\Sigma^\mathrm{inel}(\varepsilon) = \Sigma^\mathrm{total}(\varepsilon)-\Sigma^\mathrm{tr}(\varepsilon)$. For each impact electron energy $\varepsilon$, we integrate this cross section over the electron distribution $X(\varepsilon^{\prime}) f(\varepsilon^{\prime})$ to obtain excitation rates and use detailed balance to obtain relaxation rates, respecting all Pauli blocking factors and assuming that the final electrons in any three-body process carry equal energies. Since the recombination rates are much larger than the excitation rates and are purely collisional, we assign an inelastic $\nu(\omega$) at each $\omega$ to be the recombination rate of the corresponding collision excitation at $\varepsilon=\omega$. The T-matrix $\nu(\omega$) including these inelastic contributions are given by the solid red lines in Fig.\ \ref{fig:various}: they tend to modestly increase the static limit and contribute significant damping at lower frequencies, bringing the DSF into much better agreement with the reference spectrum. 

\subsection{$\nu(\omega)$ from Kubo-Greenwood}
\label{sec:kg}

Finally, we consider an independent method for calculating dynamic collision frequencies based on the work of \citet{Reinholz2000}, who relate the dynamic collision frequency $\nu^{ff}(\omega)$ to the free-free part of the dynamic conductivity $\sigma^{ff}(\omega)$ by 
\begin{equation}
  \sigma^{ff}(\omega) = Z_i n_i/[\nu^{ff}(\omega) - i\omega].
  \label{eq:reinholz}
\end{equation}

\begin{figure}
    \centering
    \includegraphics[width=0.45\textwidth]{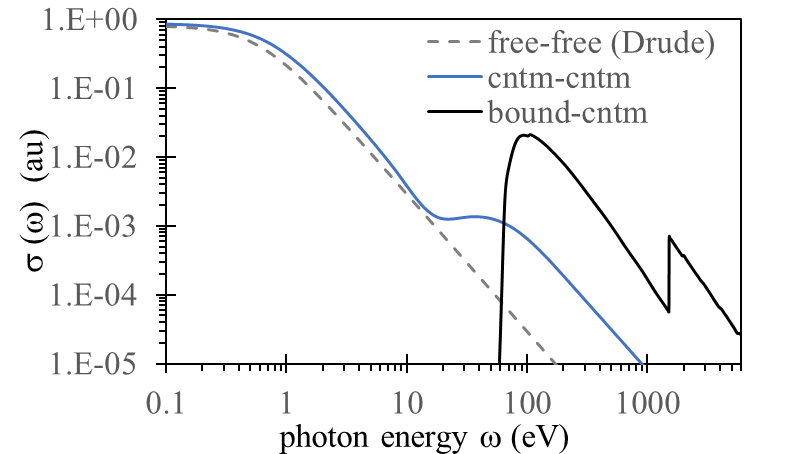}
    \caption{Dynamic conductivities for  solid-density aluminum at \SI{1}{\electronvolt} from the DFT-AA model based on the Kubo-Greenwood formalism, showing a departure of the quantum calculation from the ideal/Drude form.
    }
    \label{fig:kg}
\end{figure}

Using self-consistent data from the DFT-AA model, we implement a Kubo-Greenwood (KG) expression for the dynamic conductivity following \citet{johnson2006optical}. Figure \ref{fig:kg} shows the real part of the dynamic conductivity in the solid-density, \SI{1}{\electronvolt} aluminum case. Three curves are shown: the bound-free contribution, a Drude-like free-free contribution representing the ideal free electrons $Z_i$, and a quantum variation of the free-free contribution representing all continuum electrons. To regularize the low-frequency $1/\omega^2$ divergence, we impose a Lorentzian broadening or Drude form on the conductivity: $\sigma^{ff}(\omega) = \sigma^{ff}(0)/[1+ (\omega / \nu^{D})^2]$. 
For ideal free electrons, this Drude form is the entire free-free contribution to the dynamic conductivity. By construction, its real part perfectly satisfies the conductivity sum rule $\int_0^{\infty}  \text{Re}\{\sigma^{ff}(\omega)\} d\omega = (\pi/2) Z_i n_i$ for \emph{any} value of $\nu^{D}$, so we typically impose the Ziman value, setting $\nu^{D}=\nu^Z(0)$. The DC limit of the dynamic collision frequency in the Drude model $\nu^{ff}(0) = n_i Z_i/ \sigma^{ff}(0)$ is then identical to the regularizing frequency $\nu^{D}$, and inversion of Eq.\ \eqref{eq:reinholz} returns a constant Re\{$\nu^{ff} (\omega)$\} $= \nu^{D}$. 
\begin{figure*}
    \centering
    \includegraphics[width=0.95\textwidth]{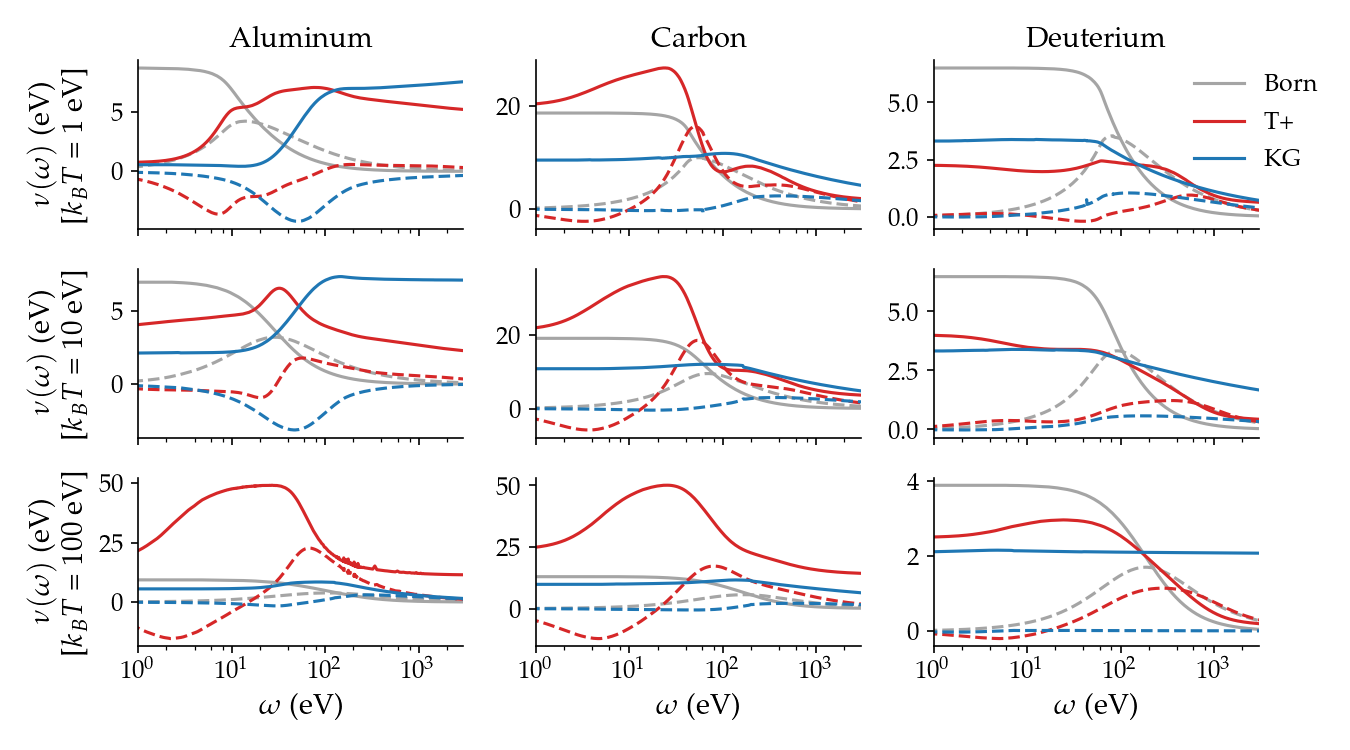}
    \caption{Real (solid) and imaginary (dashed) dynamic collision frequencies from three approaches for three materials: aluminum at a density of \SI{2.7}{\gram\per\cubic\centi\meter}; carbon at \SI{10.0}{\gram\per\cubic\centi\meter}; and deuterium at \SI{10.0}{\gram\per\cubic\centi\meter} as a function of the driving frequency $\omega$ at temperatures of \SI{1}{\electronvolt} (top row), \SI{10}{\electronvolt} (middle), and \SI{100}{\electronvolt} (bottom).} \label{fig:collisions}
\end{figure*}

Using the distorted-wave continuum electrons from the DFT-AA model in the Kubo-Greenwood expression complicates this picture. The quantum $\sigma^{cc}(\omega)$ modifies the simple $1/\omega^2$ behavior, typically adding an additional feature at low frequencies that corresponds roughly to transitions among the continuum electrons, as illustrated in Fig.\ \ref{fig:kg}. To regularize the quantum $\sigma^{cc}(\omega)$, we multiply it by a factor $1/[1+(\nu^{KG}/\omega)^2]$ and search for the collision frequency $\nu^{KG}$ that will satisfy the sum rule $\int_0^{\infty}  \text{Re} \{\sigma^{cc}(\omega)\} d\omega = (\pi/2) Z_c n_i$ \cite{faussurier2016electron}. This returns a value of $\nu^{KG}$ that is independent from the Ziman approximation, and the complex transform of $\sigma^{cc}(\omega)$ via Eq. \ref{eq:reinholz} provides an independent check on $\nu(\omega$). Here, it returns a structured collision frequency that closely resembles our T-matrix calculations below $\omega_{pl}$, as illustrated by the solid blue line in Fig.\ \ref{fig:various}. While the KG $\nu(\omega$) shifts the DSF toward the reference spectrum and reaches larger values than the T-matrix approach at higher frequencies, it does not add sufficient broadening near the peak of the DSF to be in good agreement with the reference spectrum. Still, we retain it as an independent approximation in our investigations of DSFs and stopping powers.

\subsection{$\nu(\omega)$ dependence on temperature and element}
\label{ssec:nu-dep}

Figure \ref{fig:collisions} illustrates the frequency dependence of the real (solid) and imaginary (dashed) parts of $\nu(\omega)$ for a wider range of elements and conditions. Here, and in the following sections, only three approaches to $\nu(\omega)$ are considered. ``Born'' denotes the simplest form of $\nu(\omega)$, which uses the Born-Yukawa cross section (with $Z_s=Z_c$), the ideal DOS, $S_{ii}(k)=1$, and a direct integration over $\epsilon(k,\omega)$. ``T+'' is our most consistent and complete calculation of $\nu(\omega)$, which uses the T-matrix cross section, the quantum DOS, a self-consistent $S_{ii}(k)$ (with peaks removed), the inverse (Lenard-Balescu) integration over $\epsilon^{-1}(k,\omega)$, and includes inelastic contributions to $\nu(\omega)$. ``KG'' denotes the $\nu(\omega)$ derived from the Kubo-Greenwood optical conductivities. At all temperatures and for all elements, $\nu^\mathrm{Born}(\omega)$ maintain a consistent functional form, while $\nu^{T+}(\omega)$ and, to a lesser extent, $\nu^{KG}(\omega)$ exhibit significant variations. The static limits of the three approaches, which inform the DC conductivity, also vary significantly. %
In the following sections, we explore the impact of these variations in $\nu(\omega)$ on DSFs and stopping powers.

\label{sec:results}
\section{Dynamic structure factors}
\label{sec:DSF}

X-ray Thompson scattering (XRTS) experiments have proven to be a valuable platform for benchmarking theories for WDM against experimental data~\cite{GlenzerRedmer2009, Witte2017, Sperling2015}. In XRTS experiments, hard coherent x-rays scatter from electrons in a target material, transferring some momentum $k$ and energy $\omega$. The intensity of the scattered photons is proportional to the DSF of the material, which is often expressed as a sum of three different contributions to the scattering process using the Chihara decomposition \cite{Chihara1999}:
\begin{equation}
    S(k, \omega) = |g(k)|^2 S_{ii}(k, \omega) +
    Z_0 S_{ee}(k, \omega) + S_{bf}(k, \omega).
    \label{eq:chihara}
\end{equation}
Here, the first term in the sum represents elastic scattering from tightly bound electrons that closely follow the ion motion and depends on the Fourier transform of the total electron density $g(k)$. The second term describes inelastic scattering from the $Z_0$ free electrons, and the third describes inelastic scattering from bound electrons that are photoionized in the scattering process. 

In this work, we focus only on the $Z_0 S_{ee}$ term, which is the dominant contribution to the overall scattering signal in the collective regime corresponding to smaller scattering angles or momentum transfers \cite{GlenzerRedmer2009}. It is related to the dielectric response through the fluctuation-dissipation theorem:
\begin{equation}
    S_{ee}(k, \omega) = -\frac{1}{1 - \exp(-\omega/T)}\frac{k^2}{4 \pi Z_0 n_i}\text{Im}\left[\frac{-1}{\epsilon(k, \omega)}\right].
    \label{eqn:fdt}
\end{equation}
Here, $Z_0 n_i$ is the free electron density. The dielectric function $\varepsilon(k, \omega)$ describes the linear screening response to an external driving electric field, and it can be thought of as a generalization of the dielectric constant for dynamic fields. The factor $\imag{[-1/\varepsilon]}$ is related to the energy dissipated by the system and is called the electron loss function (ELF).

In the WDM community, the free-electron dielectric response is often approximated using an RPA
dielectric function~\cite{Mahan, Arista-Brandt1984, Johnson2012, Souza2014} where the electrons are assumed to act as a uniform electron gas (UEG).
The RPA dielectric function is equivalent to the finite-temperature extension of the Lindhard dielectric function \cite{Lindhard1954}, and is accurate for non-collisional plasmas. As shown by recent XRTS experiments on warm, dense aluminum, however, the RPA does not generally provde an adequate description of the collective plasmon response in the WDM regime.
\cite{Sperling2015, Witte2017}. 

One approach to improving the RPA dielectric is to include electron correlations beyond the RPA using a local field correction (LFC). The LFC factor is introduced to account for the existence of exchange-correlation holes around the electrons, with earlier approximations for low-temperatures made by Hubbard and others (Refs.\ \onlinecite{Hubbard1958, Singwi1968} as cited in Ref.\ \onlinecite{Mahan}).
Recently, Dornheim and co-workers have developed a neural network representation for the LFC of the warm, dense UEG, trained using \textit{ab initio} path integral Monte Carlo data \cite{Dornheim2018Jan, Dornheim2018Dec, Dornheim2019, Hamann2020, dornheim2020effective, dornheim2022electronic}.

While the current progress on LFC for the UEG is promising, it does not capture distortions in the electron density that accompany the complex screening of partially ionized material in the WDM regime. The distortions lead to collisions that modify the electron response, and can be included using the Mermin ansatz\cite{Mermin1970}
\begin{equation}
    \epsilon^{\text{M}}(k, \omega) = 1 + \frac{(\omega+i\nu)[\epsilon^0(k, \omega+i\nu) - 1]}{\omega + i\nu \frac{\epsilon^0(k, \omega+i\nu) - 1}{\epsilon^0(k,0)-1}}  \;,
    \label{eq:mermin}
\end{equation}
which modifies the RPA dielectric function $\epsilon^0$ according to the collision frequencies $\nu(\omega)$.
While it is possible to combine both the effects of the LFC for correlations among the electrons and the Mermin ansatz for electron-ion collisions in an extended Mermin approach~\cite{wierling2009dynamic, Fortmann2010}, here we focus only on the impact of $\nu(\omega)$. We ensure that $S_{ee}$ satisfies the sum rule by using the ideal DOS and modified chemical potential for RPA and Born-Mermin DSFs, and by using $\mu_{AA}$ and including the factor $\xi(\varepsilon)$ for KG and T+ Mermin DSFs (see Section \ref{sec:AAnu} and Appendix \ref{appendix:DOS-change}).

\begin{figure}
    \centering
    \includegraphics[width=0.45\textwidth]{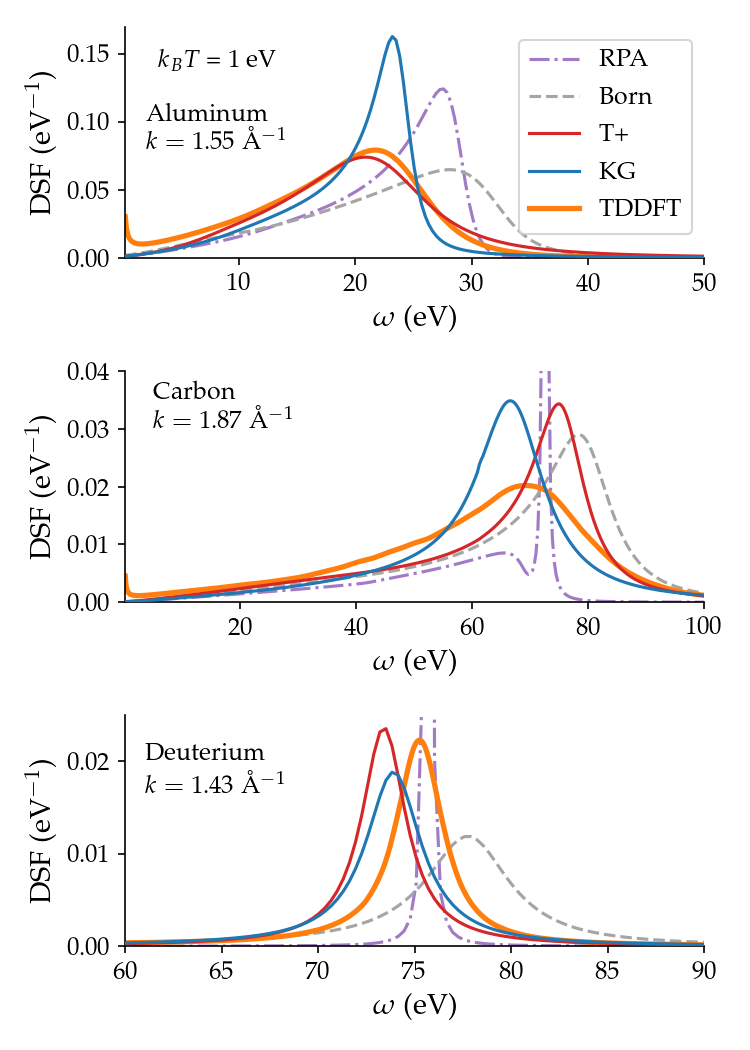}
    \caption{Dynamic structure factors for aluminum at a density of \SI{2.7}{\gram\per\cubic\centi\meter}, carbon at \SI{10.0}{\gram\per\cubic\centi\meter}, and deuterium at \SI{10.0}{\gram\per\cubic\centi\meter}. The temperature for all subplots is \SI{1}{\electronvolt} and the wavenumber for each element is specified. The inclusion of collision frequencies in the dielectric function tends to broaden and shift the DSF when compared to the RPA DSF. At these low momentum transfers, the plasmon peak for the RPA DSF for carbon and deuterium is very close to a $\delta$-function and was artificially broadened by including a small real, constant collision frequency on the order of 1 eV.}
    \label{fig:dsf-t1}
\end{figure}

\begin{figure}
    \centering
    \includegraphics[width=0.45\textwidth]{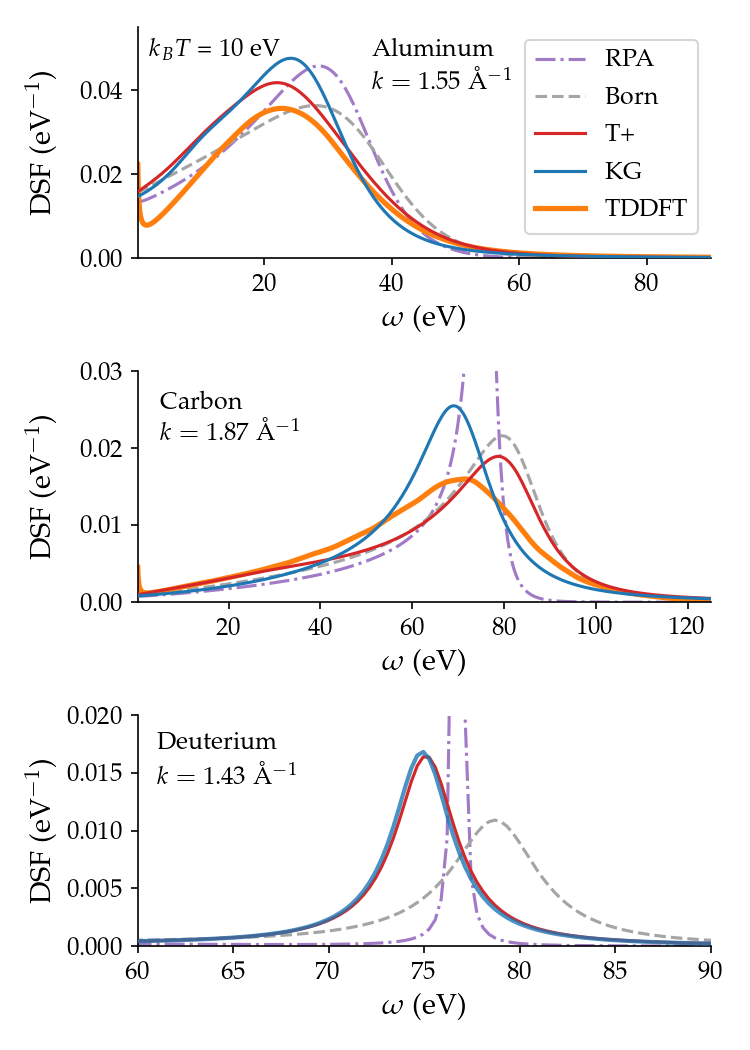}
    \caption{DSFs  for aluminum, carbon, and deuterium at the same densities and wavenumbers as in Figure \ref{fig:dsf-t1} at a temperature of $\SI{10}{\electronvolt}$. TDDFT calculations for deuterium are not included.
    }
    \label{fig:dsf-t10}
\end{figure}

Figures \ref{fig:dsf-t1} and \ref{fig:dsf-t10} show DSFs in the collective scattering regime for aluminum at \SI{2.7}{\gram\per\cubic\centi\meter}, carbon at \SI{10.0}{\gram\per\cubic\centi\meter}, and deuterium at \SI{10.0}{\gram\per\cubic\centi\meter} at thermal energies of $\SI{1}{\electronvolt}$ and $\SI{10}{\electronvolt}$. Each subplot contains DSFs constructed using the RPA dielectric (purple dash-dotted lines) and the Mermin dielectric using the Born (gray dashed lines), T+ (red solid lines), and KG (blue solid lines) collision frequencies specified in Section \ref{ssec:nu-dep}.

We find that including dynamic collision frequencies in the Mermin ansatz broadens and shifts the plasmon peaks away from the RPA DSFs \cite{Fortmann2010}. The negative imaginary values of the Born $\nu(\omega)$ shift the plasmon peaks to higher energies, while the KG and T+ collisions tend to red-shift the peaks to lower energies.

The orange lines in Figures \ref{fig:dsf-t1} and \ref{fig:dsf-t10} show DSFs calculated using TDDFT following the approach of Refs.\ \onlinecite{Baczewski2016,Baczewski2021} to directly model the electronic response to an x-ray probe. For aluminum and carbon, the free-electron contributions to the DSF are isolated by pseudizing all but the outermost valence electrons, explicitly treating only 3 and 4 electrons per atom, respectively. Deuterium is fully ionized at this density, so all electrons are explicitly treated in this case. Additional details of the TDDFT calculations are described in Appendix \ref{appendix:TDDFT}. Here, TDDFT serves as an independent and highly accurate benchmark for the approximate Mermin DSFs informed by collisions from the DFT-AA model. 

Aluminum is a nearly-free electron metal with an unstructured DOS, and is thus a good candidate for the Mermin model, which adds density corrections to the UEG picture assumed by the RPA dielectric through the relaxation-time approximation. For aluminum, both the T+ and KG collisions tend to red-shift the DSF towards the TDDFT peak, while Born collisions blue-shift the plasmon away from the TDDFT peak. The inclusion of inelastic collisions in $\nu^{T+}(\omega)$ significantly broadens the Mermin DSF, leading to good agreement with TDDFT, while the real part of $\nu^{KG}(\omega)$ does not provide significant broadening near $\omega_{pl}$ and has a much narrower plasmon peak. 

For carbon at $k_BT = \SI{1}{\electronvolt}$, the RPA DSF predicts a very narrow plasmon feature, which is expected for highly collective scattering as $k/k_c \rightarrow 0$~\cite{Mahan}, with $$k_c = \frac{n_e}{k_B T} \frac{ F_{-1/2}(\frac{\mu}{k_B T})}{ F_{1/2}(\frac{\mu}{k_B T})}$$ the inverse screening length from Ref.\ \onlinecite{GlenzerRedmer2009} and $F_j(x)$ the complete Fermi-Dirac integral of index $j$. Here, $k/k_c$ is about twice as large for aluminum as it is for carbon or deuterium in Figure \ref{fig:dsf-t1}.
The carbon in this case is also very strongly coupled ($\Gamma_{ii}=209$) and its DOS is much more structured than that of the DFT-AA model (see Figure \ref{fig:dos}). In contrast with the aluminum case, none of the approximate DSFs agree well with the TDDFT results for carbon, although all of the Mermin DSFs offer improvements over the RPA and both KG and T+ collisions offer slight improvements over Mermin with Born collisions. 

For fully-ionized deuterium at $k_BT = \SI{1}{\electronvolt}$, the RPA DSF agrees fairly well with the location of the plasmon peak predicted by TDDFT. This success can be attributed to the nearly-free electron DOS, similar to aluminum (see Figure \ref{fig:dos}). The inclusion of T+ and KG collisions results in a slightly shifted and broadened peak that is still in reasonable agreement with the TDDFT results. 

For all elements, the observed trends at $k_BT = \SI{1}{\electronvolt}$ are echoed at the higher temperature of $k_BT = \SI{10}{\electronvolt}$, as shown in Figure \ref{fig:dsf-t10}. However, thermal broadening, increasingly ideal densities of state, and weaker coupling mitigate some of the more profound differences among models. %

\section{Stopping powers}
\label{sec:dEdx}

We now consider the stopping power, a critical quantity for energy balance in inertial fusion. The stopping power is the rate of kinetic energy lost by a projectile with some initial velocity $v$ per unit length traveled through a target material and is denoted by $-dE(v)/dx $ \cite{Race2010}.

It is common to separate the stopping into contributions from collisions with the target nuclei and interactions with the target electrons. At low projectile velocities, the nuclear stopping dominates and is typically described with a classical treatment of the interactions between the projectile and target nuclei using an effective interatomic potential \cite{Wilson1977}. 
At large projectile velocities, electronic stopping dominates and must be treated in a quantum mechanical picture \cite{Schleife2015, Race2010}. Since we are interested in how electron-ion collisions affect the free electron dynamics, we ignore the nuclear stopping contribution.

Within linear response theory, the stopping power is expressed as a double integral over momentum and energy space of the electron loss function (see Appendix \ref{appendix:integrals} for a discussion of our numerical approach to computing these integrals), which is proportional to the DSF discussed in the previous section:
\begin{equation}
    -\frac{dE(v)}{dx} = \frac{2Z_p^2}{\pi v^2} \int_0^\infty \frac{dk}{k} \int_0^{kv} d\omega \omega \imag \left[ \frac{-1}{\epsilon(k, \omega)}\right].
    \label{eq:stopping}
\end{equation}
In Equation \ref{eq:stopping}, $Z_p$ and $v$ are the projectile's charge and velocity. The applicability of this formula is limited to cases where the projectile interactions with the target electrons are weak, and it has been shown to overestimate the stopping power for highly charged ions \cite{Clauser2018}.

Within TDDFT calculations, the stopping power is computed from the force exerted on the projectile by the target electrons as their response to the moving ion is simulated in real time.
The free-electron contributions to stopping powers were isolated through appropriate pseudization in the same manner as was done for the DSF calculations.
The approach closely follows previous work\cite{Magyar2016,kononov2022inprep}, and additional details are included in Appendix \ref{appendix:TDDFT}.

\begin{figure}
    \centering
    \includegraphics[width=0.45\textwidth]{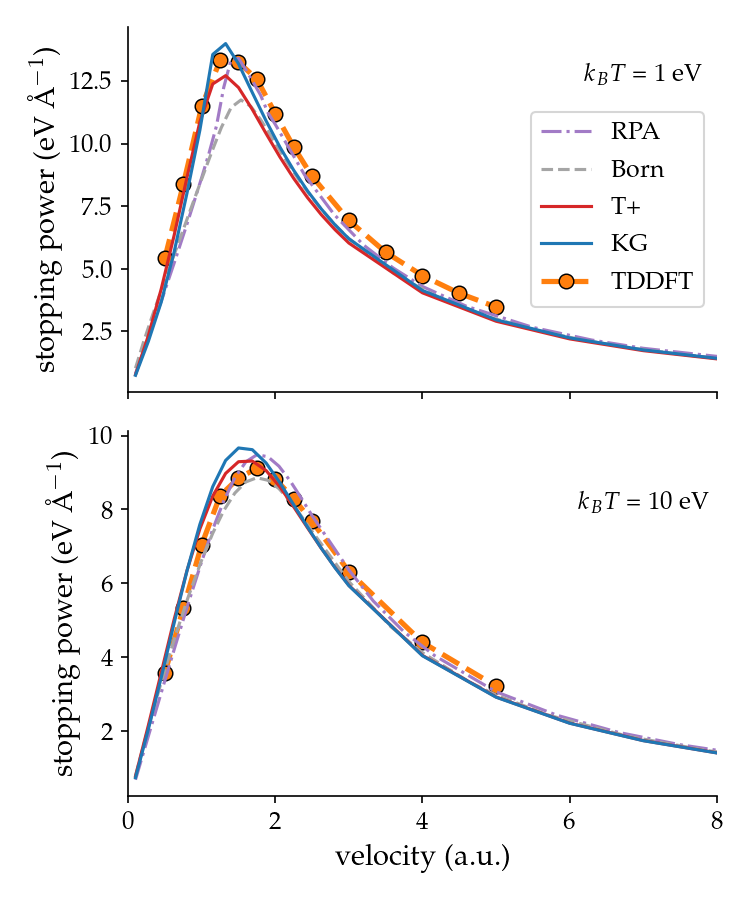}
    \caption{Electronic stopping powers for a proton traveling through aluminum at \SI{2.7}{\gram\per\cubic\centi\meter}, for temperatures of 1 and 10 \si{\electronvolt}. The lines correspond to stopping powers evaluated within the dielectric formalism using the RPA dielectric function (dot-dashed), and the Mermin dielectric function with Born, T-matrix, and KG collisions. The orange dots are stopping powers from TDDFT calculations with a 3-electron pseudo-potential.}
    \label{fig:stoppingAl}
\end{figure}

\begin{figure}
    \centering
    \includegraphics[width=0.45\textwidth]{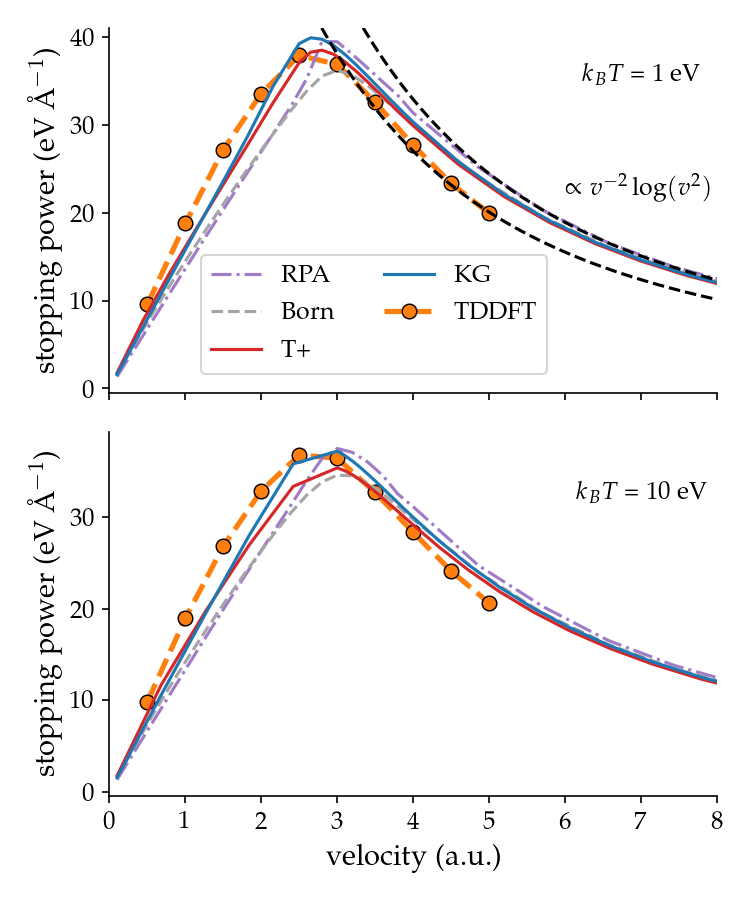}
    \caption{Electronic stopping powers for a proton traveling through carbon at \SI{10.0}{\gram\per\cubic\centi\meter}, for temperatures of 1 and 10 \si{\electronvolt}. The TD-DFT calculations use a 4-electron pseudo-potential. The dashed black lines correspond to curves proportional to the Bethe stopping formula.}
    \label{fig:stoppingC}
\end{figure}

\begin{figure}
    \centering
    \includegraphics[width=0.45\textwidth]{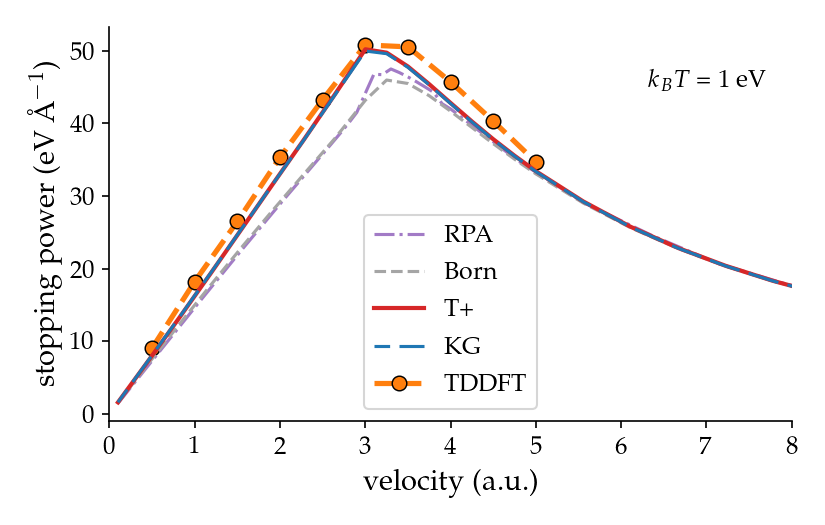}
    \caption{Electronic stopping powers for a proton traveling through deuterium with a density of \SI{10.0}{\gram\per\cubic\centi\meter} and a temperature of 1\si{\electronvolt}.}
    \label{fig:stoppingD}
\end{figure}

We evaluate the stopping power for a proton ($Z_p=1$) projectile within aluminum, carbon, and deuterium as a function of velocity in Figures \ref{fig:stoppingAl}, \ref{fig:stoppingC}, and \ref{fig:stoppingD} respectively, using the RPA and Mermin dielectric functions with the Born, T+, and KG collision frequencies discussed in Section \ref{sec:AAnu} modifying the electron loss function of Eq. \ref{eq:stopping}. These are compared to TDDFT free-electron stopping power calculations at the same conditions, which we again treat as a benchmark for the approximate model based on RPA or Mermin dielectric functions and collisions from the DFT-AA model.

For the stopping in aluminum at $k_BT = \SI{1}{\electronvolt}$ in Figure \ref{fig:stoppingAl}, the RPA stopping power agrees with the magnitude and location of the Bragg peak predicted by TDDFT but underestimates the stopping at lower velocities. The inclusion of T+ and KG collisions, which red-shifted the DSFs, significantly improves agreement with the low-velocity portion of the curve. Shifting the plasmon peak to lower energies effectively enables collective excitations by lower velocity, or lower energy, projectiles. 
By contrast, the the simple Born electron-ion collision frequency in the Mermin dielectric stopping power noticeably lowers the height of the Bragg peak and does not improve the low-velocity stopping.
At the higher temperature of \SI{10}{\electronvolt}, all of the approximate stopping powers match TDDFT quite well.

In Figure \ref{fig:stoppingC}, using T+ and KG collision frequencies in the Mermin stopping power again improves agreement with TDDFT in the low velocity regime in warm, dense carbon. At high projectile velocities, the RPA and Mermin stopping powers are larger than the TDDFT predictions; this disagreement should not be too surprising considering the disagreement found in the DSFs in Section \ref{sec:DSF}. However, the high-velocity tails for the approximate dielectric-based theories follow the Bethe stopping power formula~\cite{Bethe1930} (as cited in Ref.~\onlinecite{Correa2012}), which is proportional to $v^{-2}\log(v^2)$ as $v \rightarrow \infty$. TDDFT seems to follow the same trend with a different proportionality constant, but with only a few high-velocity data points it is difficult to determine the exact behavior. Finite-size effects and/or shortcomings of the pseudopotential approximation likely cause the TDDFT data to underestimate stopping powers at high velocities. At $\SI{10}{\electronvolt}$, the disagreement at higher velocities decreases.

Figure \ref{fig:stoppingD} shows the stopping power for deuterium with $k_BT = \SI{1}{\electronvolt}$. The agreement between the TDDFT stopping power and the Mermin stopping power using T+ and KG collisions is excellent over the whole velocity range. As discussed in the previous section, this can be attributed to the nearly-free electronic behavior in deuterium. Here, the broadening from the T+ and KG collisions seem to be necessary to improve the stopping power for low to intermediate projectile velocities, since the RPA plasmon energy agreed with TDDFT.

\section{Discussion}
\label{sec:conclusion}

We have presented several approaches for calculating dynamic collision frequencies using self-consistent quantities determined within a DFT-based average-atom model and have investigated their impact on transport and scattering properties of WDM. We find that collision cross sections based on the Born approximation cannot capture the details of strong scattering in partially ionized plasmas. Strong, fully quantum collisions have not yet been rigorously incorporated within the formalism of the generalized Boltzmann equation \cite{Reinholz2000}, so we have pursued two simple, independent approaches: first, a substitution of fully quantum T-matrix cross sections in the standard integrals for dynamic collision frequencies, and second, a transform of the complex dynamic conductivity from Kubo-Greenwood calculations. Both approaches explicitly incorporate the fully quantum density of states and account for all continuum electrons. These approaches agree well with each other in the static limit for weakly coupled plasmas, and both show radical departures from the functional form of collision frequencies based on Born cross sections. We have also explored the effects of inelastic processes on the dynamic collision frequencies.

Changes in dynamic collision frequencies $\nu(\omega)$ carry over into modifications of observable and transport properties. In particular, we use the Mermin ansatz to obtain $\nu(\omega)$-dependent dielectric functions that go beyond the RPA and directly provide DSFs and integrands for stopping powers. Using \emph{ab-initio} TDDFT as a reference model, we show that the new approaches to dynamic collision frequencies offer significant improvement over previous approaches: we find that collision frequencies computed using both Kubo-Greenwood calculations and T-matrix cross sections tend to shift plasmon features towards the TDDFT results, and that these shifts in the DSFs result in much better agreement with TDDFT stopping powers at velocities around and below the Bragg peak. This demonstrates that adequate treatments of collisions can overcome some of the shortcomings of linear-response stopping power models based on dielectric functions.

While we have not found an approach to computing $\nu(\omega)$ that gives perfect agreement with TDDFT in every case, the present results offer a computationally efficient and internally consistent approach to predicting observable and transport quantities. These improvements will increase the reliability of XRTS diagnostics and help close gaps between the computationally expensive first-principles models suitable for benchmark calculations and the computationally efficient average-atom models suitable for wide-ranging tabulation of material properties.

\section*{Acknowledgments} \label{sec:acknowledgements} 
We are grateful for conversations with Nathaniel Shaffer, Charles Starrett, Charles Seyler, and Andr\'{e} Schleife. We thank Joel Stevenson for technical support.

SBH and TWH were partially supported by the US Department of Energy, Office of Science Early Career Research Program, Office of Fusion Energy Sciences under Grant No. FWP-14-017426. TWH was also supported by the NNSA Stewardship Science Academic Programs under DOE Cooperative Agreement DE-NA0003764.
AK, AO, and ADB were supported by the US Department of Energy Science Campaign 1.
AC was partially supported by the Center for Advanced Systems Understanding (CASUS) which is financed by Germany’s Federal Ministry of Education and Research (BMBF) and by the Saxon state government out of the State budget approved by the Saxon State Parliament.
SBH, AK, AO, AC, and ADB were partially supported by Sandia National Laboratories' Laboratory Directed Research and Development Program.

Sandia National Laboratories is a multi-mission laboratory managed and operated by National Technology and Engineering Solutions of Sandia, LLC, a wholly owned subsidiary of Honeywell International, Inc., for DOE's National Nuclear Security Administration under contract DE-NA0003525.
This paper describes objective technical results and analysis.
Any subjective views or opinions that might be expressed in the paper do not necessarily represent the views of the U.S. Department of Energy or the United States Government.

\section*{Author Declarations} \label{sec:author-declarations}

\subsection*{Conflict of Interest}
The authors have no conflicts to disclose.

\subsection*{Author Contributions}
\textbf{Thomas W. Hentschel:} Conceptualization (equal); data curation (lead); formal analysis (equal); writing - original draft (lead); writing - review \& editing (equal); software (equal).
\textbf{Alina Kononov:} Conceptualization (equal); formal analysis (equal); writing - original draft (supporting); writing - review \& editing (equal).
\textbf{Alexandra Olmstead:}  formal analysis (equal);  writing - review \& editing (equal).
\textbf{Attila Cangi:} formal analysis (supporting); writing - review \& editing (equal).
\textbf{Andrew D. Baczewski:} Conceptualization (equal); formal analysis (equal); writing - original draft (supporting); writing - review \& editing (equal).
\textbf{Stephanie B. Hansen:} Conceptualization (equal); formal analysis (equal); writing - original draft (supporting); writing - review \& editing (equal); software (equal).

\appendix
\section{Using the quantum DOS in the RPA dielectric}
\label{appendix:DOS-change}
The RPA dielectric function $\epsilon^0$ is written in terms of the density response for the non-interacting electron gas $\chi^0$ \cite{Mahan}:
\begin{equation*}
    \epsilon^0(\mathbf{k}, \omega) = 1 - \frac{4\pi}{k^2}\chi^0 (\mathbf{k}, \omega),
\end{equation*}
where
\begin{equation*}
    \chi^0 (\mathbf{k}, \omega) = -2\int \frac{d\mathbf{p}}{(2\pi)^3}
    \frac{f(\varepsilon(\mathbf{p}+\mathbf{k})) - f(\varepsilon(\mathbf{p}))}{\varepsilon(\mathbf{p}+\mathbf{k}) - \varepsilon(\mathbf{p}) - (\omega + i\delta)}
\end{equation*}
and where $\delta \rightarrow 0^+$, $f(\varepsilon)$ is the Fermi-Dirac occupation factor (Eq.\ \eqref{eq:FD-distro}), and
$\varepsilon(\mathbf{p}) = p^2/2$.
By using spherical coordinates for $\mathbf{p}$, the density response can be written as
\begin{align*}
    \chi^0 (\mathbf{k}, \omega) &= \int_0^\infty \frac{p^2 dp}{\pi^2}F(\varepsilon(p); k, \omega) \\
    &= \int_0^\infty d\varepsilon X^i(\varepsilon) F(\varepsilon; k, \omega)
\end{align*}
with
\begin{multline*}
F(\varepsilon; k, \omega) = - \frac{1}{2k \sqrt{2\varepsilon}}f(\varepsilon) ~ \times \\
 \Bigg[ \log\left( \frac{k^2/2 + k \sqrt{2\varepsilon} - (\omega + i\delta ) }{k^2/2 - k \sqrt{2\varepsilon } - (\omega + i\delta ) }\right)+ \\
 \log\left( \frac{k^2/2 + k \sqrt{2\varepsilon} + (\omega + i\delta) }{k^2/2 - k \sqrt{2\varepsilon} + (\omega + i \delta )  }\right)  \Bigg] 
\end{multline*}
and $X^i(\varepsilon)$ the ideal DOS.

We propose swapping $X^i(\varepsilon)$ with the quantum DOS obtained from the DFT-AA model (or, equivalently, multiplying the integrand by the ratio of the quantum DOS to the ideal DOS $\xi(\varepsilon) = X(\varepsilon)/X^i(\varepsilon))$. This modification replaces the quadratic energy spectrum of ideal continuum states embedded in $F(\varepsilon;k, \omega)$ with the quantum behavior predicted by the DFT-AA model, allowing us to use the DFT-AA chemical potential while still fulfilling sum rules and accounting for all of the continuum electrons.

A similar procedure can be carried out to incorporate $\xi(\varepsilon)$ in the Ziman collision frequency (Eq.\ \eqref{eq:Ziman}) \cite{StarrettTab}.

\section{Numerically calculating the dielectric stopping power integrals}
\label{appendix:integrals}

\begin{figure}
    \centering
    \includegraphics[width=0.43\textwidth]{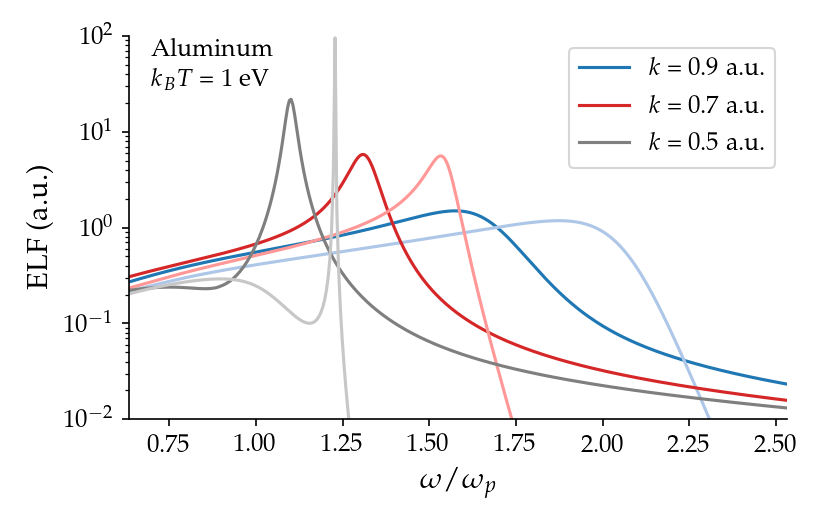}
    \caption{Electron loss function (ELF) as a function of $\omega$ for various wavenumbers. The lighter-colored lines correspond to the RPA dielectric, while the darker-colored lines correspond to the Mermin dielectric using KG+ electron-ion collision frequencies. As $k \rightarrow 0$, the ELF using the RPA dielectric develops a sharp feature that makes numerical integration difficult. }
    \label{fig:elf}
\end{figure}

For finite temperatures, the double integral in the stopping power formula (Eq.\ \eqref{eq:stopping}) must be computed numerically, where the dielectric function is also determined from a numerically evaluated integral. To efficiently compute these integrals, we examine the integrands themselves. Figure \ref{fig:elf} illustrates the electron loss function (ELF) $ \imag (-1 /\epsilon(k, \omega))$ of aluminum with a density of \SI{2.7}{\gram\per\cubic\centi\meter} at a temperature of \SI{1}{\electronvolt} as a function of $\omega$ for various values of $k$. As $k$ is decreased, the ELF using the RPA dielectric function (lighter-colored lines) develops a sharp feature resembling a $\delta$ function at the plasmon excitation energy predicted for small $k$ \cite{Mahan}. By contrast, the ELF with the Mermin dielectric (darker-colored lines) converges to a broadened peak as $k \rightarrow 0$, due to the electron-ion collisions (here, we use the KG+ collision frequencies discussed in the main text for illustration). This semi-log plot also reveals qualitatively different behavior for energies greater than the plasmon peak: while the RPA ELF abruptly falls to zero, the Mermin ELF decays much more slowly.

Integrating over the Mermin ELF is fairly straightforward numerically, but the RPA ELF is more difficult because of the sharp feature at small $k$. To resolve this difficulty, we first find the location of the peak and use this information to perform the inner, $\omega$-integral. Although approximate dispersion relations for the plasmon peak exist, these are typically only accurate for certain regions of phase space (i.e. weakly degenerate electron gas) \cite{Thiele2008}. Instead, we find the plasmon peak numerically, first using a bisection method applied directly to the ELF and resorting to a secant method for finding the zeros of the real part of the dielectric function (which corresponds to the ELF peak for small $k$ \cite{Mahan}). When the width of the peak falls below our minimum integration grid spacing as $k \rightarrow 0$, we replace the ELF peak with an exact $\delta$-function at the peak position, weighted to ensure that the $f$-sum rule is satisfied (Eq.\ \eqref{eq:sumrule}).

\begin{figure}
    \centering
    \includegraphics[width=0.48\textwidth]{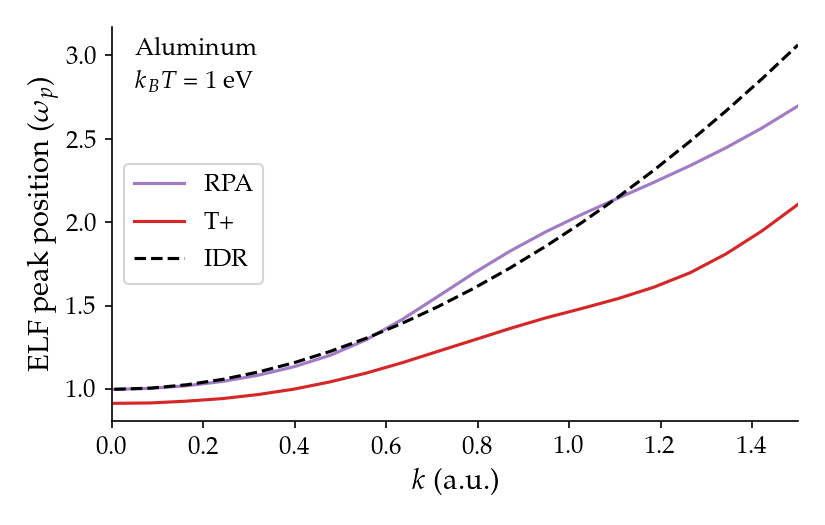}
    \caption{ELF dispersion as a function of the wavenumber. The dispersion curves for the RPA ELF and Mermin ELF using T+ collisions are calculated numerically, while the IDR curve is an approximate analytical function from \cite{Thiele2008}. In this example, the IDR curve agrees fairly well with the RPA curve for $k < 0.4$~a.u.
    }
    \label{fig:dispersion}
\end{figure}

Our numerically calculated dispersion relations for aluminum are shown in Figure \ref{fig:dispersion}. We compare our results to the improved dispersion relation (IDR) from work by Thiele and others that is accurate for a larger range of wavenumbers and degeneracies than the typical Gross-Bohm dispersion relation  (see Equations 21 \& 22 in  Ref.~\onlinecite{Thiele2008}).

To certify the accuracy of the $\omega$-integral for the RPA and Mermin ELF, we compare our results against the $f$-sum rule

\begin{multline}
    \int_0^{kv} d\omega \omega \imag \left[ \frac{-1}{\epsilon(k, \omega)}\right] + \int_{kv}^{\infty} d\omega \omega \imag \left[ \frac{-1}{\epsilon(k, \omega)}\right] \\
    = \int_0^{\infty} d\omega \omega \imag \left[ \frac{-1}{\epsilon(k, \omega)}\right] = \frac{\pi}{2}\omega_p^2
    \label{eq:sumrule}
\end{multline}
where $\omega_p$ is the plasma frequency. The first term is the integral used in the stopping power calculations, which must return the constant $\pi\omega_p^2/2$ when added to the second term (which we also compute). If this equality is satisfied, we trust the results of our quadrature scheme since the integrand is always non-negative. In this work, we accept the quadrature results if the sum rule is satisfied to within 5\%. If it is not satisfied, we can increase the accuracy of our integrator.

\begin{figure}[h]
    \centering
    \includegraphics[width=0.48\textwidth]{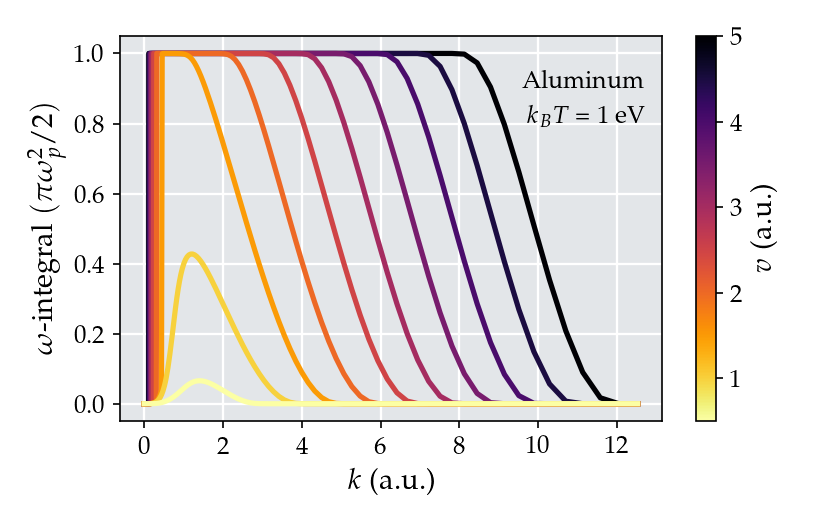}
    \caption{The $\omega$-integral for aluminum at \SI{1}{\electronvolt} as a function of $k$ for various velocities $v$ (indicated by the color bar), using the RPA dielectric function. For a given velocity, the points on the curve are checked against the sum rule for each value of $k$. 
    }
    \label{fig:kintegrand}
\end{figure}

The numerical results for the $\omega$-integral ($\int_0^{kv}d\omega \omega \imag(-1/\epsilon)$) are shown in Figure \ref{fig:kintegrand} as a function of $k$ for various values of the velocity $v$.
For a given element, temperature, and density, the overall structure of these functions depends on the velocity: for small values of $k$, the upper limit of the $\omega$-integral $kv$ exceeds the dispersion of the plasmon peak. If $v$ is large enough, after some value of $k$ the majority of the plasmon peak is contained in the integration range $[0,kv]$, and the curve for the $\omega$-integral flattens off at the sum rule value. As $k$ increases, the dispersion of plasmon surpasses $kv$. Eventually, the integration range will contain essentially none of the plasmon peak, and the integral will be 0. 

Notice that for the RPA ELF, the rise up to the sum rule is abrupt since we approximate the RPA ELF as a $\delta$-function once the width of the peak falls below our minimum integration grid spacing, which occurs for small $k$.
Finally, these curves are multiplied by $2Z_1^2/(\pi v^2) \times 1/k$ and integrated over $k$ to obtain the stopping power as a function of $v$. Since the functional form is relatively well behaved, simple quadrature approaches can be used.

\section{Details of the real-time time-dependent density functional theory calculations}
\label{appendix:TDDFT}

DSF and stopping power calculations were performed using an implementation of real-time TDDFT first reported in Refs.~\onlinecite{Magyar2016} and \onlinecite{Baczewski2016}, respectively.
This implementation is an extension of the Vienna \emph{ab initio} simulation package~\cite{kresse1996efficient,kresse1996efficiency,kresse1999ultrasoft}, which relies on the projector augmented-wave (PAW) formalism~\cite{blochl1994projector} to accurately and efficiently represent the electron-ion interaction.
In both types of calculations, the central equations of motion are the time-dependent Kohn-Sham (TDKS) equations, with initial conditions determined by the self-consistent orbitals from a Mermin-Kohn-Sham DFT calculation~\cite{kohn1965self,mermin1965thermal}.
These orbitals are computed for an atomic configuration either corresponding to a known crystal structure or sampled from a Born-Oppenheimer molecular dynamics trajectory~\cite{mattsson2004designing}.
To represent the effects of electronic temperature, the Mermin-Kohn-Sham orbitals are occupied according to the Fermi-Dirac distribution at a fixed temperature and self-consistently determined chemical potential.
These occupations are held constant over the course of the integration of the TDKS equations.
All calculations use the adiabatic local density approximation \cite{zangwill:1980,zangwill:1981}, in which the time-dependent exchange-correlation potential is approximated %
to have only local spatial and temporal dependence on the electronic charge density.

The TDKS equations are numerically integrated using the Crank-Nicolson method~\cite{qian2006time}. 
The discrete update equations are solved using %
the generalized minimal residual (GMRES)~\cite{saad1986gmres} iterative algorithm with a default residual of $10^{-12}$.
Due to the ionic motion in the stopping power calculations, it is necessary to add a gauge correction to the TDKS equations to properly account for the time dependence of the projectors and preserve unitarity of the integration (and thus probability and charge conservation).
To this end, we have followed the approach in Ref.~\onlinecite{ojanpera2012nonadiabatic} in our implementation.

The DSF calculations rely on a real-time implementation of a linear response calculation, similar to the one that was first reported in Ref.~\onlinecite{sakko2010time} and later refined and adapted to WDM in Ref.~\onlinecite{Baczewski2016}.
Aluminum calculations at a temperature of \SI{1}{\electronvolt} (\SI{10}{\electronvolt}) contained 64 (32) atoms and 5 (64) electronic bands per atom with a 3-electron pseudopotential, plane-wave cutoff energy of \SI{500}{\electronvolt}, and reciprocal space integration using a $4\times 4\times 4$ $\Gamma$-centered quadrature.
Carbon calculations contained 128 atoms and 3 (12) electronic bands per atom for a temperature of \SI{1}{\electronvolt} (\SI{10}{\electronvolt}) with a 4-electron pseudopotential, plane-wave cutoff energy of \SI{1000}{\electronvolt}, and reciprocal space integration using the Baldereschi mean-value point \cite{baldereschi:1973}.
Deuterium calculations contained 256 atoms and 1 electronic bands per atom for a temperature of \SI{1}{\electronvolt} with an all-electron pseudopotential, plane-wave cutoff energy of \SI{2800}{\electronvolt}, and reciprocal space integration using a $7\times 7\times 7$ $\Gamma$-centered quadrature.
All the DSF calculations used a time step of \SI{1}{\atto\second}.

Stopping powers were computed from time-dependent Hellmann-Feynman forces acting on a proton moving through the supercell.
Details regarding the choice of proton trajectory and stopping power extraction are documented in Ref.\ \onlinecite{kononov2022inprep}.
Aluminum calculations contained 256 atoms with a plane-wave cutoff of \SI{750}{\electronvolt}.
Deuterium calculations contained 1728 atoms 
with a plane-wave cutoff energy of \SI{2000}{\electronvolt}.
All the stopping power calculations sampled reciprocal space using the $\Gamma$ point only.
For fast protons traversing aluminum or carbon with $v\geq 1.5$a.u., the time step scaled inversely with proton velocity to maintain a fixed displacement of about \SI{0.02}{\angstrom} within each step.
Slower protons required shorter time steps of 0.3\,--\,\SI{0.4}{\atto\second} to achieve convergence.
Similar time steps of 0.2\,--\,\SI{0.4}{\atto\second} were used for deuterium calculations.
All other parameters remained the same as listed for the DSF calculations above.

\bibliography{stopping.bib}

\end{document}